\documentclass[aps,prx, amsfonts, superscriptaddress, amssymb, amsmath, reprint, showkeys, nofootinbib, raggedbottom, longbibliography]{revtex4-1}
\usepackage[english]{babel}
\usepackage[utf8]{inputenc}
\usepackage[colorinlistoftodos, color=green!40, prependcaption]{todonotes}
\usepackage{amsthm}
\usepackage{mathtools}
\usepackage{physics}
\usepackage{xcolor}
\usepackage{graphicx}
\usepackage{adjustbox}
\usepackage{placeins}
\usepackage[T1]{fontenc}
\usepackage{lipsum}
\usepackage{csquotes}
\usepackage{float}
\usepackage{afterpage}
\usepackage{booktabs}
\usepackage{bm}
\usepackage[version=4]{mhchem}

\usepackage[pdftex, pdftitle={Article}, pdfauthor={Author}]{hyperref} % For hyperlinks in the PDF
\graphicspath{ {Figures/} }

\begin{document}
\title{Single-\textit{q} Cycloid and Double-\textit{q} Vortex Lattices in Layered Magnetic Semimetal \ce{EuAg4Sb2}}

\author{Paul M. Neves}
\affiliation{Department of Physics, Massachusetts Institute of Technology, Cambridge, MA 02139, USA}
\author{Takashi Kurumaji}
\affiliation{Division of Physics, Mathematics and Astronomy, California Institute of Technology, Pasadena, California 91125, USA}
\author{Joshua P. Wakefield}
\affiliation{Department of Physics, Massachusetts Institute of Technology, Cambridge, MA 02139, USA}
\author{Arno Hiess}
\affiliation{Institut Laue Langevin, 71 Avenue des Martyrs, F-38000 Grenoble Cedex 9, France}
\affiliation{European Spallation Source ERIC, P.O. Box 176, 22100 Lund, Sweden}
% https://orcid.org/0000-0002-6457-691X
\author{Paul Steffens}
\affiliation{Institut Laue Langevin, 71 Avenue des Martyrs, F-38000 Grenoble Cedex 9, France}
\author{Navid Qureshi}
\affiliation{Institut Laue Langevin, 71 Avenue des Martyrs, F-38000 Grenoble Cedex 9, France}
\author{Robert Cubitt}
\affiliation{Institut Laue Langevin, 71 Avenue des Martyrs, F-38000 Grenoble Cedex 9, France}
\author{Lisa M DeBeer-Schmitt}
\affiliation{Large Scale Structures Section, Neutron Scattering Division, Oak Ridge National Laboratory, Oak Ridge, TN, USA}
\author{Johanna C. Palmstrom}
\affiliation{National High Magnetic Field Laboratory, LANL, Los Alamos, NM, USA}
\author{Satoru Hayami}
\affiliation{Graduate School of Science, Hokkaido University, Sapporo 060-0810, Japan}
\author{Marek Bartkowiak}
\affiliation{PSI Center for Neutron and Muon Sciences, CH-5232 Villigen PSI, Switzerland.}
\author{Markus Zolliker}
\affiliation{PSI Center for Neutron and Muon Sciences, CH-5232 Villigen PSI, Switzerland.}
\author{Jonathan S. White}
\affiliation{PSI Center for Neutron and Muon Sciences, CH-5232 Villigen PSI, Switzerland.}
\author{Joseph G. Checkelsky}
\email{checkelsky@mit.edu}
\affiliation{Department of Physics, Massachusetts Institute of Technology, Cambridge, MA 02139, USA}

\date{\today} % Leave empty to omit a date

\begin{abstract}
Recently, a host of exotic magnetic textures such as topologically protected skyrmion lattices has been discovered in several bulk metallic lanthanide compounds. In addition to hosting skyrmion phases, a hallmark of this class of materials is the appearance of numerous spin textures characterized by a superposition of multi-$q$ magnetic modulations: spin moir\'{e} superlattices. The nuanced energy landscape thus motivates detailed studies to understand the underlying interactions. Here, we comprehensively characterize and model the three zero-field magnetic textures present in one such material, EuAg$_4$Sb$_2$. Systematic symmetry breaking experiments using magnetic field and strain determine that the ground state incommensurate magnetic phase (ICM1) is single-$q$. In contrast, ICM2 and ICM3 are both double-$q$, \textit{i.e.}, spin moir\'{e} superlattices. Further, through application of polarized small angle neutron scattering and spherical neutron polarimetry, we demonstrate that ICM1 is a single-$q$ cycloid and ICM2 and ICM3 are double-$q$ vortex lattices, with Eu moments lying in the $ab$-plane in zero field and with a ferromagnetic component at finite field. Despite the quasi-2D nature of EuAg$_4$Sb$_2$, the modulations propagate out of the \textit{ab}-plane, leading to a shift of the spin texture between triangle lattice planes. Further, the ICM3 to ICM2 transition includes an unusual 45$^\circ$ rotation of the magnetic vortex lattice. Motivated by the coexistence of such drastically different phases in this compound, we conclude by developing a phenomenological model to understand the stability of these states. Our experimental probes and theoretical modeling definitively characterize three different and tunable phases in one material, and provide insight for the design of new topological spin-texture materials.
\end{abstract}
\maketitle

\section{Introduction}
Density wave instabilities are ubiquitous in quantum materials and are often intertwined with other emergent phenomena of interest \cite{gruner1988dynamics, gruner2000density}. Key examples include charge density waves (CDWs) and pair density waves (PDWs) in cuprates \cite{agterberg2020physics, hayden2024charge}, spin density waves in iron-based superconductors \cite{stewart2011superconductivity}, CDWs in transition metal dichalcogenides (TMDs) \cite{manzeli20172d}, and CDWs in kagome metals \cite{tan2021charge, teng2023magnetism}. Generally, the presence of such modulations represents a connection between the electronic structure and other degrees of freedom of the system. However, despite the prevalence of such density wave instabilities, a complete understanding of the exact roots of their stability, and a-priori prediction of such instabilities remain a challenge.
% For example, the relative importance of Fermi surface nesting remains a topic of debate \cite{johannes2008fermi}, while detailed comparison between the bandstructure and density wave often reveal subtle connections between the density wave instability and the Fermi surface. Conventional wisdom suggests that SDWs are common in lanthanide metals due to the Ruderman–Kittel–Kasuya–Yosida (RKKY) interaction and details of the Fermi surface \cite{freeman1972energy}. As such, new materials platforms which support such phenomena and incisive probes thereof are thus highly sought to examine this interplay in detail.
%the rich incommensurate phases in the electronically clean \ce{EuAg4Sb2} offer an excellent platform to examine such concepts in more detail.
% centrosymmetric Gd-containing \ce{Gd2PdSi3} \cite{kurumaji2019skyrmion}, \ce{GdRu2Si2} \cite{khanh2020nanometric}, \ce{Gd3Ru4Al12} \cite{hirschberger2019skyrmion}, and non-centrosymmetric Eu-containing \ce{EuAl4} \cite{Takag:2022} and \ce{EuNiGe3} \cite{Singh:2023}

A variety of multi-\textit{q} incommensurate magnetic phases (ICMs) have recently attracted interest in quasi-2D lanthanide materials where both frustration and conduction electron mediated interactions are important \cite{wang2020skyrmion}. This combination is believed to drive the formation of a multitude of ICM phases, including vortex lattices and skyrmion lattices \cite{ozawa2016vortex, ozawa2017zero, hayami2021noncoplanar}. A common observation in this material class is a remarkable number of ICM phases of similar energy, suggesting a nuanced competition of relevant interactions. Additionally, such materials often exhibit a robust coupling between electrical transport and magnetic order---a property especially relevant to potential technological applications in spintronics. We note that vortex lattices, familiar from studies of %represents a topologically nontrivial magnetic phase, which has famously attracted attention from ideas about 2D systems and
the Berezinskii–Kosterlitz–Thouless transition \cite{kosterlitz1973ordering} and superconducting vortex lattices \cite{blatter1994vortices, abrikosov2004nobel}, are a natural manifestation of this in quasi-2D systems.

\begin{figure*}[htb]
	\includegraphics[width=\textwidth]{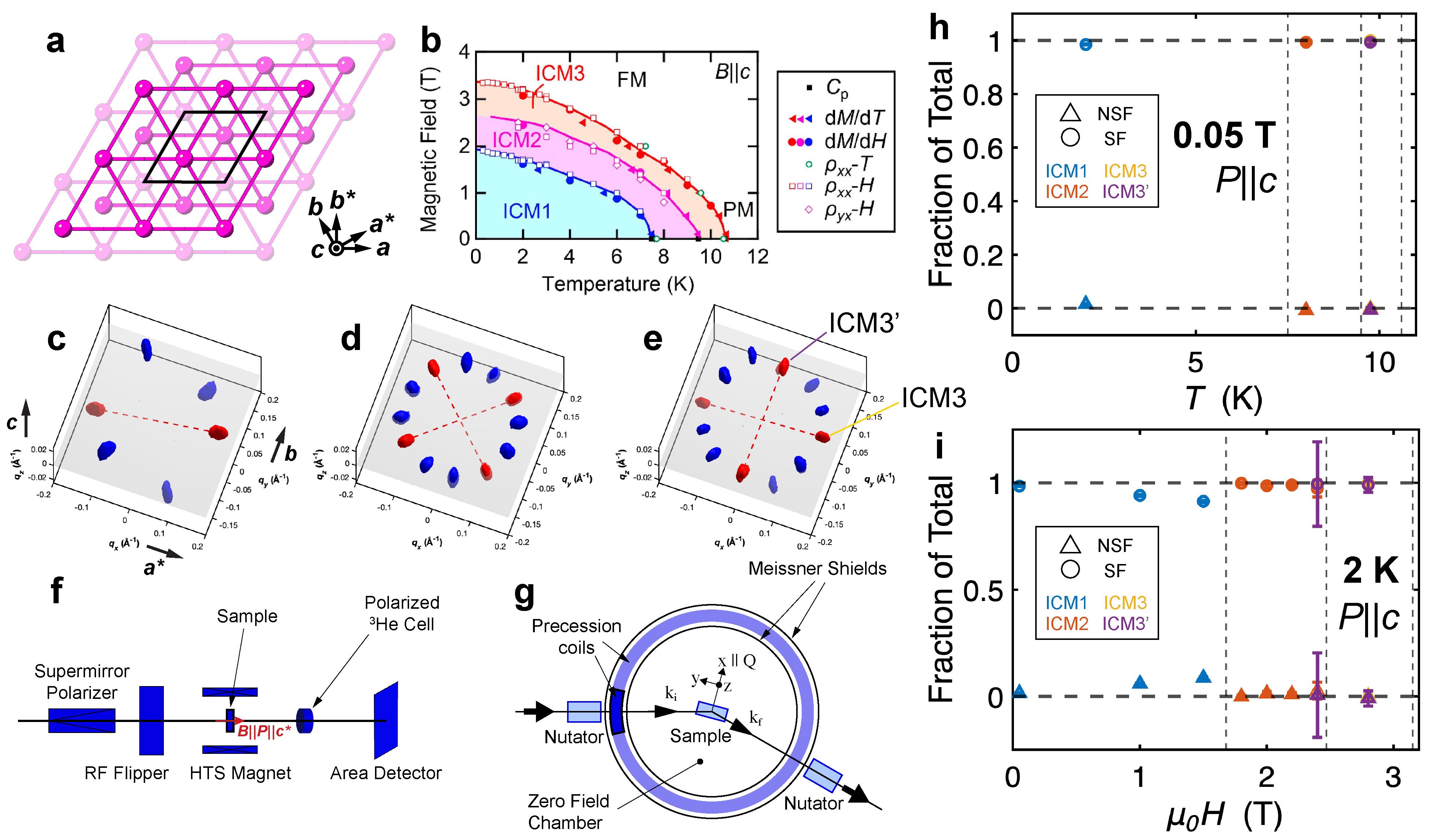}
	\caption{\label{fig:pSANS} \textbf{a} Rhombohedral ABC stacked europium triangle lattice layers viewed along the $c$-axis. The unit cell is indicated with a black outline.
    \textbf{b} $H||c$ phase diagram for \ce{EuAg4Sb2} containing the three incommensurate phases ICM1-3, from \cite{kurumaji2025electronic}. Legend indicates the measurement from which each indicated point was derived.
    \textbf{c-e} 3D SANS isosurface plots of ICM1-3 highlighting one magnetic domain each in red. The data are taken in zero field at 2.1 K, 8 K, and 9.5 K, respectively. The data are inversion symmetrized and has a high temperature background subtracted.
    \textbf{f} Schematic of the polarization-analysis setup used in the SANS experiments.
    \textbf{g} A schematic of Cryopad used to perform spherical neutron polarimetry (SNP) measurements. Note the conventional coordinate system used in polarization measurements: $\bm{x}$ is along the scattering vector $\bm{Q}$, $\bm{z}$ is the vertical axis in the instrument, and $\bm{y}$ completes the orthogonal coordinate system.
    \textbf{h} Polarization-dependent intensity of peaks in 50 mT of field along the $c$-axis as a function of temperature in different phases.
    \textbf{i} Polarization-dependent intensity of peaks at 2 K for various fields along the $c$-axis in different phases. In \textbf{h,i}, the data are background subtracted and corrected for imperfect polarization and analysis efficiency. SF is indicated with circles while NSF is indicated with triangles. Error bars are the 95\% confidence interval from a fit of the peak intensity. ICM3 (yellow) corresponds to the $H0L$-type peak while ICM3' (purple) corresponds to the $HHL$-type peak (see \textbf{e}).}
\end{figure*}

We examine here \ce{EuAg4Sb2}, a rhombohedral, centrosymmetric (spacegroup No. 166, $R\overline{3}m$), quasi-2D material hosting a Eu-triangle lattice with a Eu-Eu distance of 4.7 \AA\ (see Fig. \ref{fig:pSANS}\textbf{a}). The Eu layers are separated by 8.2 \AA\ with a \ce{Ag2Sb} layer which contributes the conduction electrons. \ce{EuAg4Sb2} orders at $T_\mathrm{N}=10.7$ K, and exhibits three ICM phases upon cooling which are all also accessible with applied magnetic field $\bm{H}||\bm{c}$ (see Fig. \ref{fig:pSANS}\textbf{b-e}) \cite{kurumaji2025electronic}. In this paper, we demonstrate through strain- and field-control of magnetic domains and polarized neutron measurements that ICM1 is a single-\textit{q} cycloid and ICM2-3 are two inequivalent double-\textit{q} vortex lattices. The propagation vectors of each phase are out of the $ab$-plane by up to 5 degrees. This means that the spin texture shifts coherently from one triangle layer of europium to the next. Further, the high symmetry of the parent compound and low symmetry of the magnetic propagation vectors lead to a large number of configurational and inversion domains, providing a rich set of addressable states and domain walls that may host distinct functionalities, a potentially advantageous property for spintronics applications \cite{catalan2012domain, sharma2017nonvolatile}. These vortex lattice states open a new platform for studying topologically nontrivial spin textures in metals, with possible applications in spintronic devices through strain or field control \cite{haykal2020antiferromagnetic, sukhanov2019giant, liyanage2024skyrmion}.

Compared to other recently discovered double-\textit{q} vortex phase materials such as \ce{Sr3Fe2O7} \cite{andriushin2024reentrant}, these phases in \ce{EuAg4Sb2} display a remarkable coupling with the electronic properties, with ICM2 and ICM3 significantly enhancing the magnetoresistance and concomitant decrease in the Hall conductivity \cite{kurumaji2025electronic, green2025mapping, malick2024large, green2025robust}. This is understood to result from a resonant condition of $2k_f\sim q$ and $J_K\sim E_F$ (where $k_F$ is the Fermi momentum, $q$ is the magnetic propagation vector, $J_K$ is the Kondo coupling strength, and $E_F$ is the Fermi energy) combined with a electron mean free path much longer than the spin modulation period. This leads to a coherent renormalization of the band structure to a new larger magnetic unit cell, becoming a so-called spin moir\'{e} superlattice \cite{shimizu_spin_2021, kurumaji2025electronic}. This has the effect of opening gaps which enhance the mass of the bands, thus altering the transport properties as observed. The opening of these gaps can also be understood as the energetic driver of the stability of these phases \cite{solenov2012chirality}. Further, the limited transport response in ICM1 can now also be understood as partly originating from the limited ability of a single-\textit{q} modulation to gap the Fermi surface compared to a multi-\textit{q} spin texture during mini-band formation. Further, such spin moir\'{e} engineering offers a path to the realization of the quantum topological hall effect \cite{hamamoto_quantized_2015}. Such behavior motivates the full characterization of each phase, which we pursue herein.

The close proximity and ability to tune between distinct ICM multi-\textit{q} phases in this material also %. The deep connection between the electronic and magnetic properties, as well as the rich and nuanced energy landscape
motivates further characterization of their character and their stability mechanisms. Our phenomenological model points to a complex energy landscape which enables the transition between such different phases. Anisotropy and a small four-spin interaction are also found to be important to understanding the phase diagram in \ce{EuAg4Sb2}. Such insights advance our understanding of the engineering and control of topological spin textures in spin moir\'{e} superlattice materials, and our demonstration of new tunability such as strain control represents an exciting avenue for the manipulation of multi-\textit{q} spin textures in technologically relevant materials. The vortex lattice states present in \ce{EuAg4Sb2}, analogous to other BKT or superconducting lattices and distinct from skyrmion lattices, make it a model platform for studying tunable topological textures in a spin moir\'{e} system.

\section{Solving the Magnetic Structures}
Any spin texture in a material with one magnetic site per unit cell (as is the case in \ce{EuAg4Sb2}) can be described as
\begin{equation} \label{eq:1}
    \bm{M}(\bm{r})=\sum_i \bm{M}_i \exp \left( \text{i} \bm{q}_i \cdot \bm{r} \right)
\end{equation}
where $i$ is the index of the magnetic propagation vector $\bm{q}_i$, $\bm{r}$ is the real-space position, $\bm{M}(\bm{r})$ is the real-space magnetization density, and $\bm{M}_i$ is the complex Fourier amplitude which characterizes the phase and magnitude of the $i^{th}$ superimposed spin wave. Uniform magnetization can be included with a $\bm{q}=0$ component. In order to determine the complete nature of such magnetic textures, it is necessary to answer two key questions. First, which magnetic diffraction peaks belong to which magnetic domains, \textit{i.e.}, which $\bm{q}_i$ should be included in Eq. \ref{eq:1}. For example, a single-$q$ texture would be striped, while a double-$q$ texture would be a square, rectangular, rhombic, or parallelogram tiling (see Fig. \ref{fig:strain}\textbf{a-b}). Second, one must determine the spin components associated with each propagation vector, that is, what $\bm{M}_i$ to use in Eq. \ref{eq:1}. A variety of potential spin structures are depicted in Fig. \ref{fig:strain}\textbf{c}, demonstrating the rich variety of potential spin textures one must distinguish between.

\subsection{Determination of Domains and multi-$q$ Nature}
\begin{figure*}[htb]
	\includegraphics[width=\textwidth]{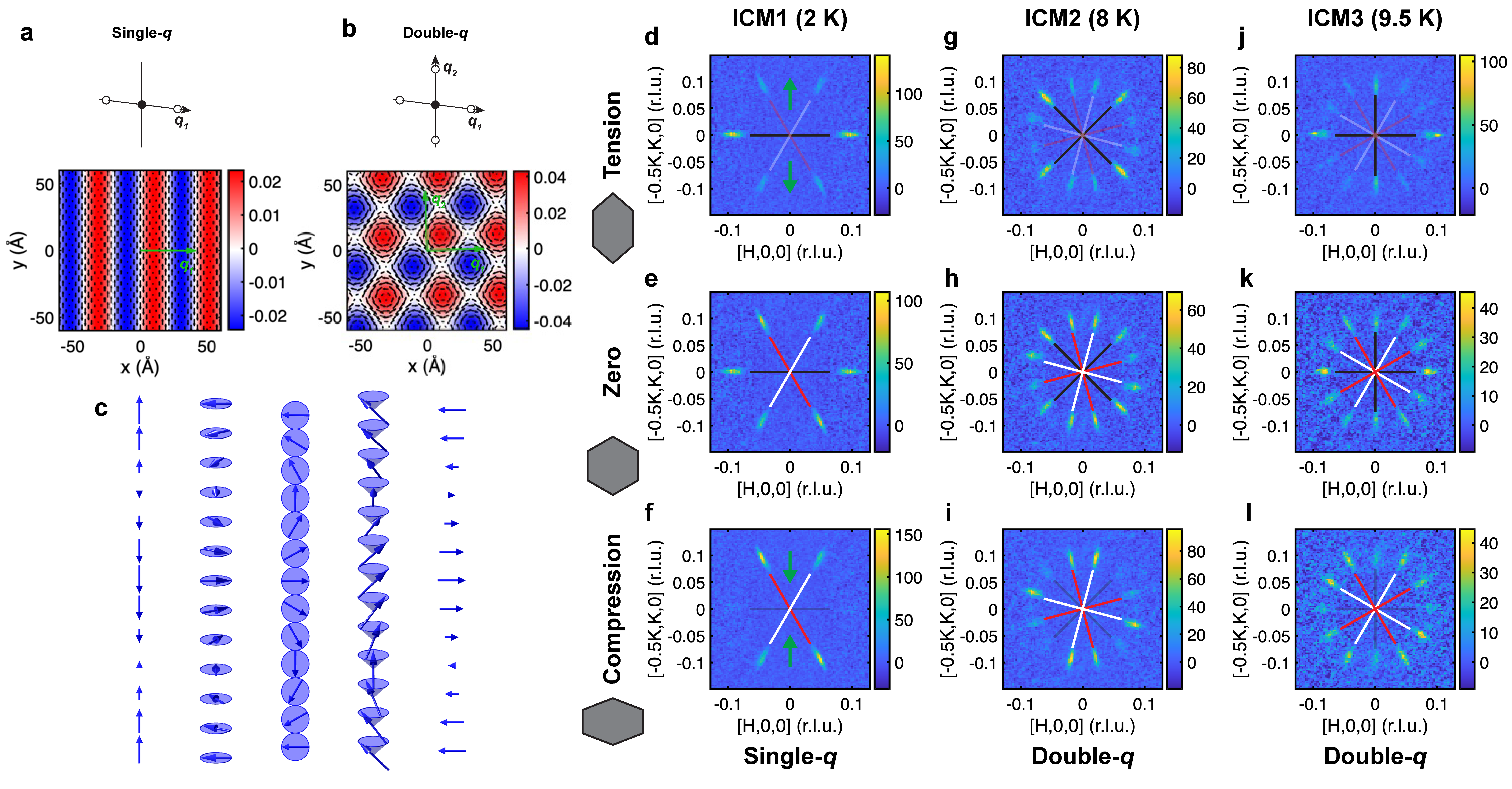}
	\caption{\label{fig:strain} \textbf{a-b} single-$q$ and double-$q$ order for a transverse modulated spin structure. The color-scale is the curl (vorticity) of the texture.
    \textbf{c} Various possible spin textures associated with one spin modulation. From left to right: longitudinal, proper screw spiral, cycloid, conical, and transverse spin structures.
    \textbf{d,g,h} SANS pattern measured under tension along the vertical direction (see green arrows in \textbf{d}) at 2 K in ICM1, 8 K in ICM2, and 9.5 K in ICM3.
    \textbf{e,h,k} SANS pattern measured with no applied strain (after the crystal broke) under the same conditions.
    \textbf{f,i,l} SANS pattern measured under uniaxial compression along the vertical axis (see green arrows in \textbf{f}).
    A cartoon (not to scale) of the strained crystal is shown with gray hexagons.
    The domain structure is indicated with red, white, and black lines. Crossed lines indicate a domain of double-$q$ states.
    SANS data are inversion symmetrized, smoothed with a half pixel gaussian kernel, and has a 15 K background measurement subtracted.}
\end{figure*}

A concise demonstration of the magnetic domain structure is possible by cooling the sample with applied uniaxial strain, with additional evidence presented in supplementary section I. The unstrained crystal has a uniform population of domains (see Fig. \ref{fig:strain} \textbf{e,h,k}). Upon application of tension (elongating the crystal along the [-1,2,0] direction, see methods), one pair of peaks becomes dominant in ICM1, while two sets of peaks become dominant in ICM2 and ICM3 (see Fig. \ref{fig:strain} \textbf{d,g,j}). The application of compression favors the peaks which become weaker under tension (see Fig. \ref{fig:strain} \textbf{f,i,l}). This confirms ICM1 is single-$q$, while ICM2 and ICM3 are double-$q$. ICM2 has nearly perpendicular propagation vectors forming a rhombic lattice, while ICM3 has orthogonal (and symmetry-distinct) propagation vectors forming a rectangular lattice. One domain of peaks in ICM1-3 is depicted in red in Fig. \ref{fig:pSANS}\textbf{c-e}, visualized in 3D. Here, we focus on the magnetic domains of one crystallographic twin. Further details of the connection between crystallographic and magnetic domains are found in supplementary information.

With the modest applied strain (no more than $\sim$0.25\%), the domains could be controlled by strain-cooling, though with more strain it may be possible to switch domains directly within the ordered state. This strain-control of magnetic domains suggests that there must be some magneto-elastic coupling that changes the relative energies of the different magnetic domains upon the application of strain. The interplay of magnetism, domains, and strain has been reported previously in several compounds \cite{ramazanoglu2011giant, dhital2012effect, nii2015uniaxial, park2018effect, kim2020controlling}.  The small residual intensity in the minority peaks under tension or compression are likely either due to insufficient or inhomogeneous application of strain, or from scattering of the unstrained portion of the crystal held by glue to the jaws.

\subsection{Polarized Neutron Scattering Measurements}
To determine the microscopic aspects of the structure we turn to polarized neutron measurements. The general experimental setup and coordinate system are depicted in Fig. \ref{fig:pSANS} \textbf{f,g}. The ratio of spin-flip (SF) and non-spin-flip (NSF) scattering when the polarization axis is aligned with the $c$-axis is depicted in Fig. \ref{fig:pSANS} \textbf{h,i}. ICM2-3 are essentially entirely spin flip, indicating that the spin component of the magnetic moment is perpendicular to the polarization axis (the $c$-axis). Therefore, the spins lie in the $ab$-plane to within experimental resolution, excluding any phases with finite skyrmion number. The finite magnetization with applied field originates fully from a uniform ($\bm{q}=0$) moment. ICM1 also predominantly lies in the $ab$-plane, but does begin to gain a spin component out of the plane as field is applied along the $c$-axis, from $1.5\pm0.2\%$ at $\mu_0H=0.05$ T and $T=2$ K up to $8.7\pm0.5\%$ spin flip scattering at $\mu_0H=1.5$ T and $T=2$ K.

Polarized  small angle neutron scattering (SANS) indicates that the majority of the spin component is within the $ab$-plane, and has at least a strong component perpendicular to the magnetic propagation vector $\bm{q}$. However, any component of spin along $\bm{q}$ (as would be present in a longitudinal spin modulation or cycloid) would not be measurable in SANS, as in SANS the scattering vector $\bm{Q}$ is equal to the magnetic propagation vector $\bm{q}$.

\begin{figure*}[htb]
    \includegraphics[width=\textwidth]{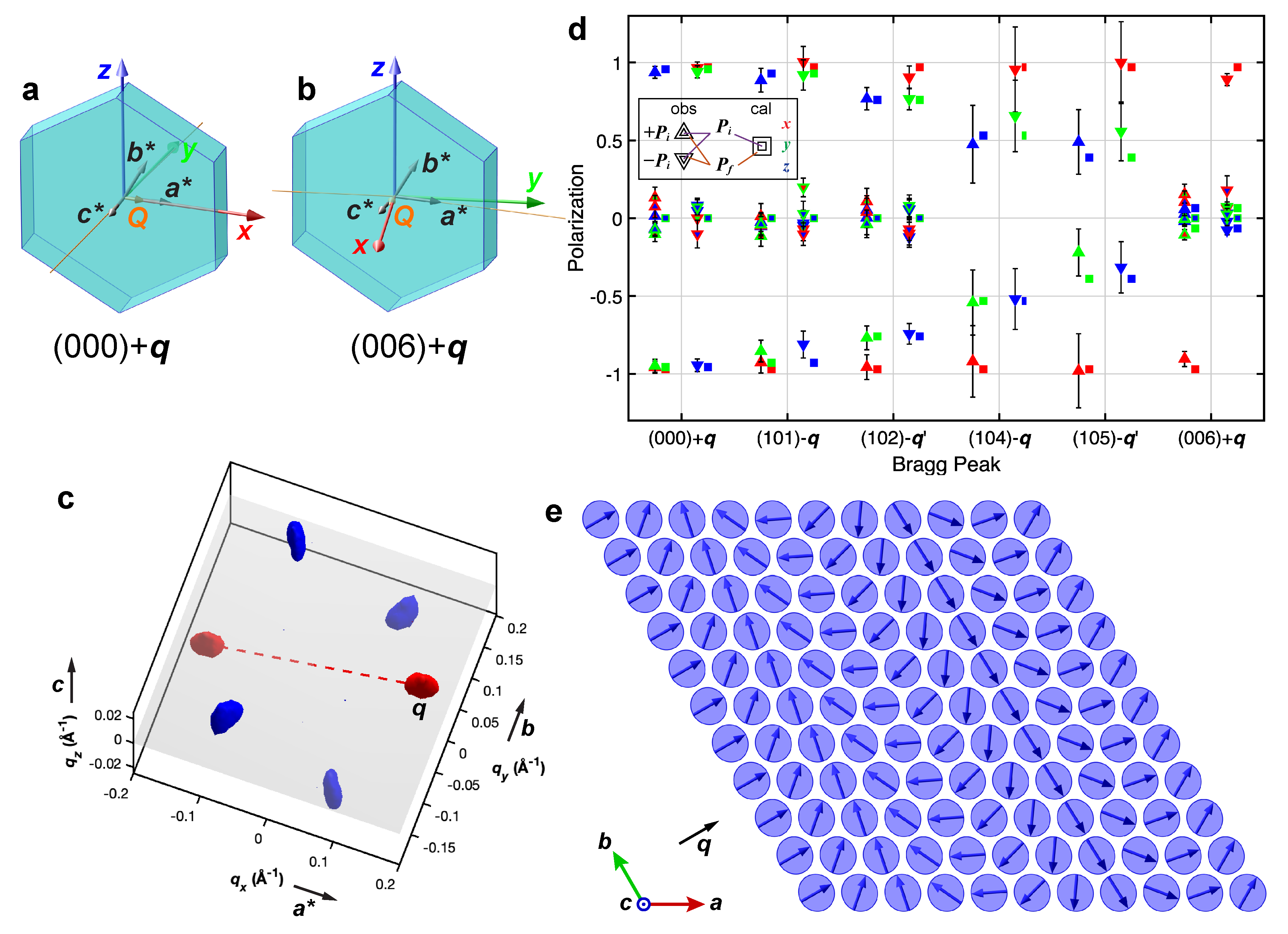}
    \caption[SNP Model of ICM1]{\textbf{a-b} Geometry of the crystal and neutron spin polarization analysis axes for the (000) and (006) satellite peaks. The coordinate axes of the polarization are depicted with red, green, and blue arrows, and the scattered neutron path is indicated in yellow. The yellow arrow corresponding to the scattering vector $\bm{Q}$.
    \textbf{c} 3D SANS intensity of ICM1 for reference, with one domain highlighted in red.
    \textbf{d} The polarization matrix elements for different magnetic satellite peaks and the corresponding model intensities from the model depicted in \textbf{e}. The incident and outgoing neutron polarizations are indicated with the inner and outer marker color, respectively. Red, green, and blue correspond to $x$, $y$, and $z$, respectively. The upwards and downwards triangles indicate a positive and negative incident neutron polarization, respectively. Measured matrix elements are indicated with triangles, while the elements calculated from the model are indicated with squares. All errorbars include both statistical and estimated systematic uncertainties.
    \textbf{e} Schematic spin structure of ICM1. A single Eu-layer is shown for simplicity. Propagation direction is also depicted.
    }
    \label{fig:ICM1_snp}
\end{figure*}

To probe the longitudinal component of ICM1-3, we perform spherical neutron polarimetry (SNP) on the primary propagation vectors in the zeroth zone and on magnetic satellite peaks of the nuclear Bragg peaks. In particular, measurement of the (006) satellite allows a clear probe of the spin component along $\bm{q}$ as the scattering vector of the (006)$\pm \bm{q}$ peak is nearly perpendicular to $\bm{q}$ (see Fig. \ref{fig:ICM1_snp}\textbf{a,b}).

\begin{figure*}[htb]
\includegraphics[width=\textwidth]{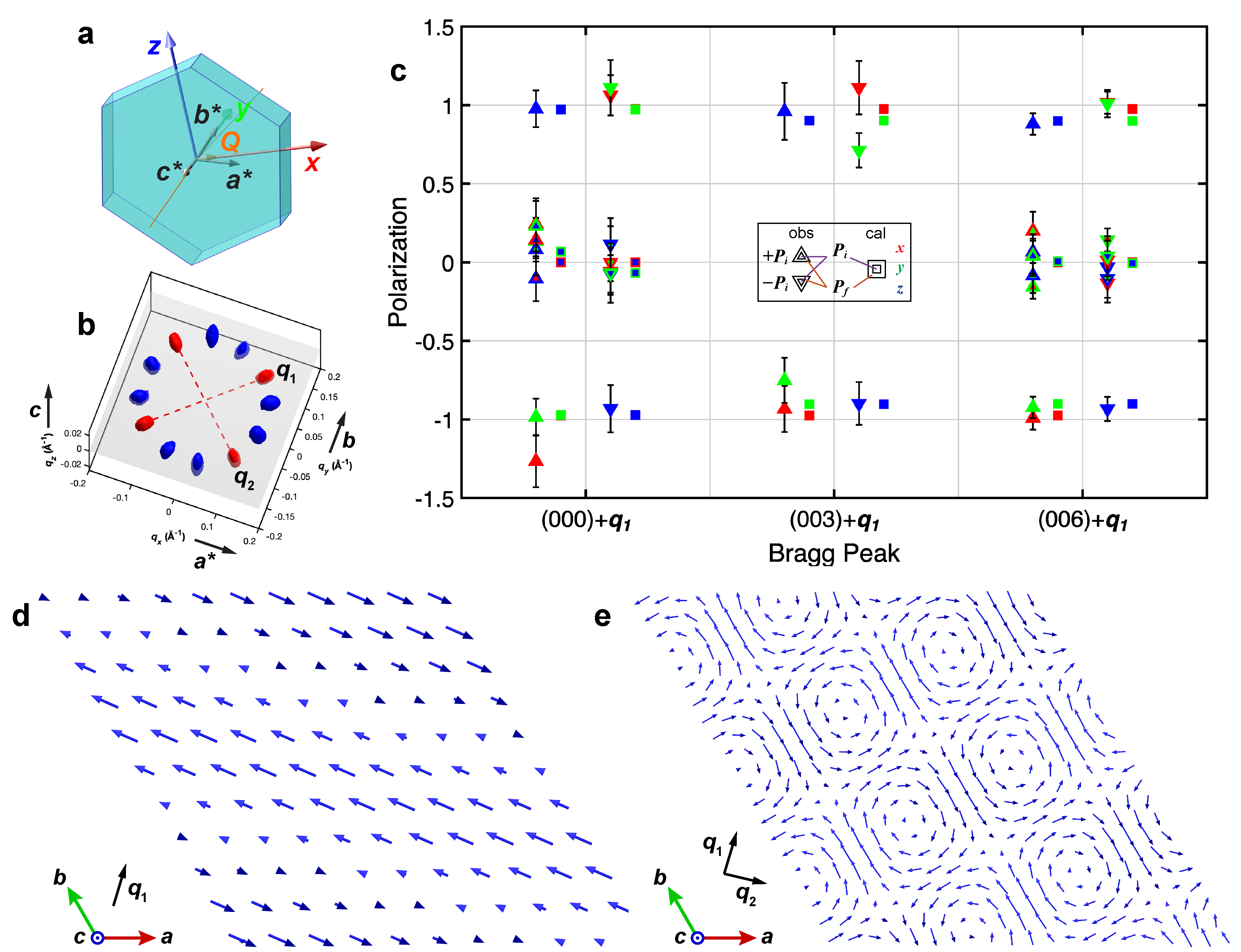}
    \caption[SNP Model of ICM2]{\textbf{a} Geometry of the crystal and polarization axes for the (000) satellite peak. The coordinate axes of the polarization are depicted with red, green, and blue arrows, and the scattered neutron path is indicated in yellow. The yellow arrow corresponding to the scattering vector $\bm{Q}$. \textbf{b} 3D SANS intensity of ICM2 for reference, with one domain highlighted in red. \textbf{c} The polarization matrix elements for different magnetic satellite peaks and the corresponding model intensities from the model depicted in \textbf{d}.  The incident and outgoing neutron polarizations are indicated with the inner and outer marker color, respectively. Red, green, and blue correspond to $x$, $y$, and $z$, respectively. The upwards and downwards triangles indicate a positive and negative incident neutron polarization, respectively. Measured matrix elements are indicated with triangles, while the elements calculated from the model are indicated with squares. All errorbars include both statistical and estimated systematic uncertainties.
    \textbf{e} The sum of the pattern depicted in \textbf{d} superimposed with the corresponding pattern for the second ICM2 propagation vector.}
    \label{fig:ICM2_snp}
\end{figure*}

For each ICM phase, the polarization matrix (see methods for more details) is measured for the symmetry-independent magnetic satellites of the nuclear Bragg peaks. The polarization matrix and model are depicted for each ICM phase in Fig. \ref{fig:ICM1_snp}\textbf{d}, \ref{fig:ICM2_snp}\textbf{c}, and \ref{fig:ICM3_snp}\textbf{a,b}. Further details of the models and fitting are presented in Supplementary Materials. We find that ICM1 is best modeled as a single-\textit{q} cycloid where the Eu moment rotates in the \textit{ab}-plane (see Fig. \ref{fig:ICM1_snp}\textbf{e}). Our model indicates a slight ellipsoidal character where the minor axis along $b$ is 94(1)\% as strong as the major axis along $a^*$ (see supplemental information), which is consistent with the proposal of Ref.~\cite{green2025robust}. ICM2 is best modeled as a rhombohedral double-\textit{q} vortex lattice, where two transverse spin modulations (see Fig. \ref{fig:ICM2_snp}\textbf{d}) are superimposed to create the vortex lattice in the \textit{ab}-plane (see Fig. \ref{fig:ICM2_snp}\textbf{e}). ICM3 is also found to be a double-\textit{q} rectangular vortex lattice, where two symmetry distinct transverse spin modulations (see Fig. \ref{fig:ICM3_snp}\textbf{c,d}) superimpose to create the vortex lattice (see Fig. \ref{fig:ICM3_snp}\textbf{e}). No significant spin component out of the \textit{ab}-plane was detected to within experimental uncertainty in any of the phases. ICM1 approximately satisfies the saturated moment condition, however the (anti-)vortex cores of ICM2-3 do not. To reconcile this, at high temperature, the cores may be either disordered or fluctuating, and at low temperature and high field, these cores field-polarize. This effect would produce a small $2q_1$-type peak intensity in field, which is also observed experimentally (see supplemental information).
% domains highlighted in red
% needed to rotate crystal (refer to teal hexagons)

\begin{figure*}[htb]
\includegraphics[width=\textwidth]{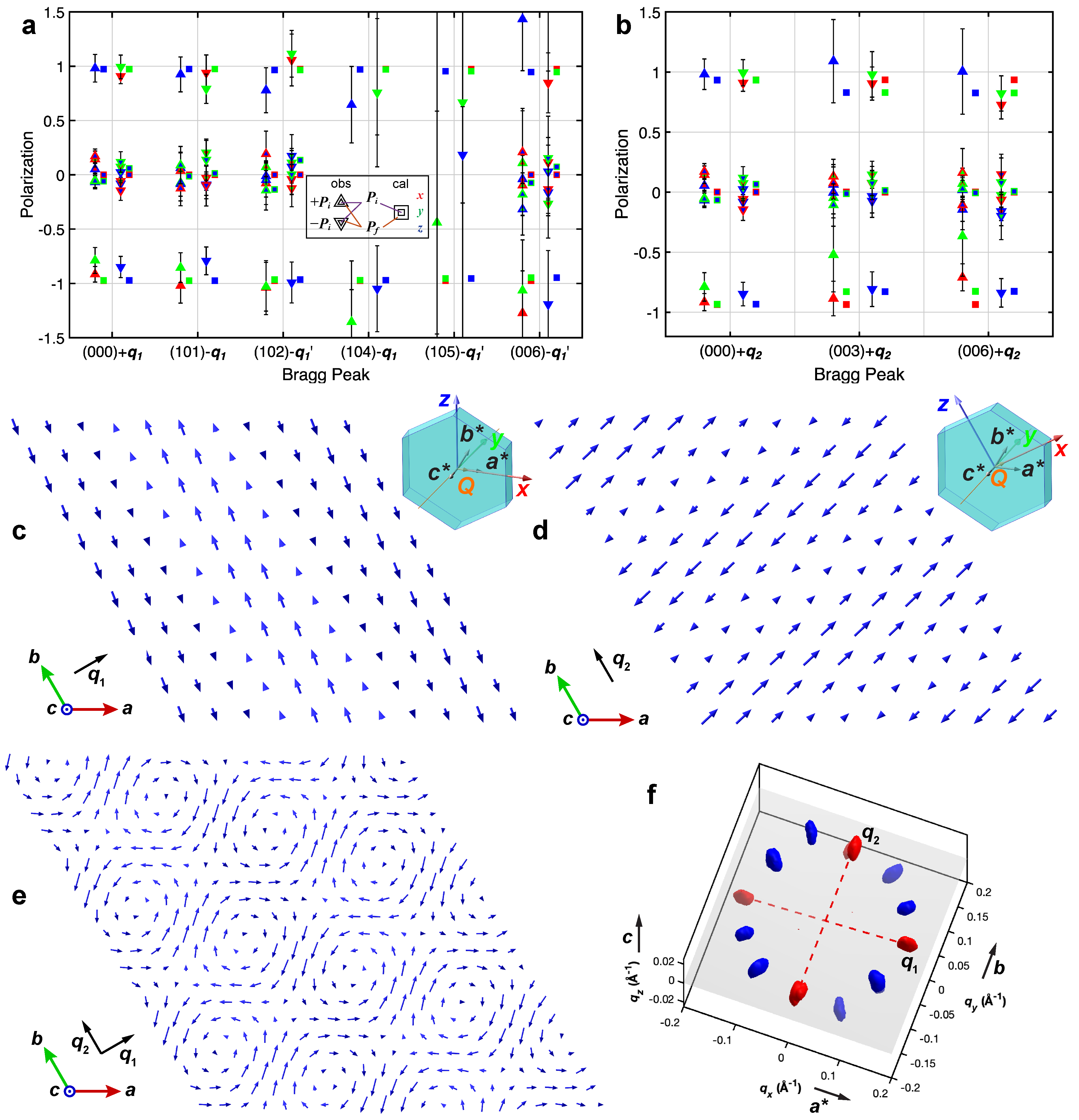}
    \caption[SNP Model of ICM3]{\textbf{a-b} The polarization matrix elements for different magnetic satellite peaks of the \textbf{a} $(H0L)$ and \textbf{b} $(HH0)$ type, and the corresponding model intensities from the models depicted in \textbf{c} and \textbf{d}, respectively. The incident and outgoing neutron polarizations are indicated with the inner and outer marker color, respectively. Red, green, and blue correspond to $x$, $y$, and $z$, respectively. The upwards and downwards triangles indicate a positive and negative incident neutron polarization, respectively. Measured matrix elements are indicated with triangles, while the elements calculated from the model are indicated with squares. All error-bars include both statistical and estimated systematic uncertainties.
    Inset of \textbf{c,d} depict the geometry of the crystal and polarization axes for the (000) satellite peak. The coordinate axes of the polarization are depicted with red, green, and blue arrows, and the scattered neutron path is indicated in yellow. The yellow arrow corresponding to the scattering vector $\bm{Q}$. \textbf{e} The superposition of the patterns depicted in \textbf{c-d}, representing the complete ICM3 texture. \textbf{f} 3D SANS intensity of ICM3 for reference, with one domain highlighted in red.}
    \label{fig:ICM3_snp}
\end{figure*}

\section{Phenomenological Model}
\begin{figure*}[htb]
    \centering
    \includegraphics[width=\textwidth]{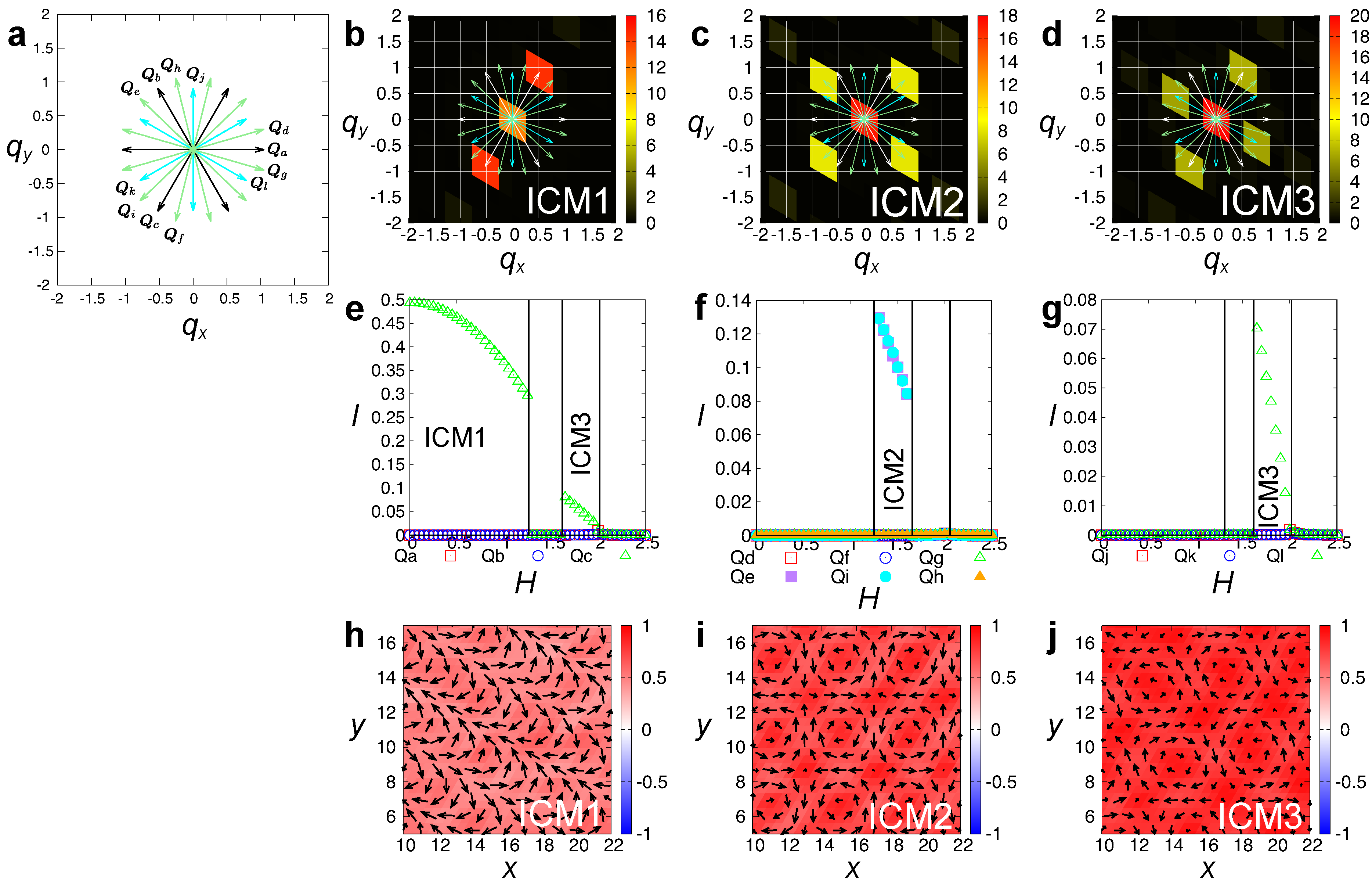}
    \caption[Phenomenological Momentum Space Modeling]{\textbf{a} Propagation vectors considered in model (see Methods). Black (white in \textbf{b-d}) arrows correspond to the ICM1 (and ICM3) vectors, green arrows correspond to the ICM2 vectors, and blue arrows correspond to the ICM3 vectors. \textbf{b-d} The propagation vectors corresponding to ICM1-3. \textbf{e-g} The model peak intensity for each of the three peak types as a function of applied magnetic field out of plane which reproduces experimental results. \textbf{h-j} Real-space model spin structure in ICM1-3. The in-plane spin is depicted with an arrow, and the $z$ component is indicated with red.}
    \label{hayami}
\end{figure*}

Finally, we present a phenomenological model for the magnetism in \ce{EuAg4Sb2} constructed in momentum space which reproduces the observed ICM phases. Such models have been applied to related systems to model phase diagrams with incommensurate order \cite{hayami2024stabilization}. A generalized momentum-space Hamiltonian is constructed as
\begin{equation}
    H = -2 J \sum_{\nu,\alpha,\beta} \Gamma^{\alpha \beta}_\nu S^\alpha_{\bm{Q}_\nu} S^\beta_{-\bm{Q}_\nu} - H \sum_i S^z_i + H_{4}
\end{equation}
where $J$ is an overall energy scale, $\Gamma$ is a generalized anisotropic spin interaction in momentum space, $S$ is the Fourier transformed spin moment, $\alpha$ and $\beta$ are the cartesian directions, $\nu$ is the index of the propagation vector (see included propagation vectors in Fig. \ref{hayami}\textbf{a-d}), and $i$ is a site index. Generally, changing which $\Gamma^{\alpha\beta}$ elements are non-zero will shift the preferred ordering between \textit{e.g.} a transverse spin modulation, a cycloid, and a helix. The second term in the Hamiltonian is the Zeeman energy term.

With the first two terms, it is possible to stabilize either ICM1 and ICM3, or ICM1 and ICM2, but not all three phases together. The introduction of a four-spin interaction term $H_4$ of the form
\begin{equation}
    H_4 = - \frac{B}{N} (\bm{S}_{\bm{Q}_{e}} \cdot \bm{S}_{-\bm{Q}_{e}}) (\bm{S}_{\bm{Q}_{i}} \cdot \bm{S}_{-\bm{Q}_{i}})
\end{equation}
where $\bm{Q}_{e}$ and $\bm{Q}_{i}$ are the two propagation vectors of ICM2 (see Fig. \ref{hayami}\textbf{c}), and N represents the total number of spins. We also consider the symmetry-related four-spin interactions. This allows the creation of a model which stabilizes all three phases at different applied magnetic fields.

To stabilize all three phases, we consider two components of $\Gamma$: an $xy$-type component $G$ in $\Gamma^{xx}$ and $\Gamma^{yy}$, and a transverse type which favors spin perpendicular to the ordering wave vector
\begin{equation}
    \Gamma_{\bm{Q}} = \begin{pmatrix}
        G+A\cos(2\theta) & -A\sin(2\theta) & 0\\
        -A\sin(2\theta) & G-A\cos(2\theta) & 0\\
        0 & 0 & 0\\
        \end{pmatrix}
\end{equation}
where $\theta$ is the angle made between $\bm{Q}$ and the $x$ axis. The absence of significant out-of-plane moment observed with polarized neutron measurements justifies the easy-plane anisotropy in $\Gamma_{\bm{Q}}$. The value of $\Gamma_{\bm{Q}}$ is in general a continuous function of $\bm{Q}$, but we only consider it at the discrete $\bm{Q}$ vectors approximately associated with ICM1-3 in this model, as depicted in Fig. \ref{hayami}\textbf{a-d}, and the $B$ term is only included connecting the ICM2 vectors. Tuning the values of $A$, $G$, and $B$ for each of the vectors to $G_1=1$, $G_2=0.95$, $G_3=0.9$, $A_1=0.05$, $A_2=0.10$, $A_3=0.16$, and $B=0.09$ successfully stabilized the observed ICM1-3 textures (see Fig. \ref{hayami}\textbf{e-j}).

The strength of $\Gamma$ at all three propagation vectors is similar, which matches with the experimentally observed close proximity of all three phases being tunable with field or temperature. Additionally, a four-spin interaction is indeed necessary to explain the presence of all three phases, as is the case in other systems \cite{forgan1989observation, ozawa2016vortex, hayami2017effective, paul2020role}, though interestingly some combinations of ICM1-3 can still be stabilized without the four-spin interaction. This four-spin interaction could perhaps originate from the itinerant nature of the present system, or the saturation moment condition. One key insight into the stability of the double-$q$ ICM2 and ICM3 states as opposed to a single-$q$ version is that the single-$q$ transverse spin modulations would deviate more from the saturated moment condition than the vortex phases. Simulated annealing at finite temperature stabilizes the ICM3 state, suggesting that ICM3 tends to be more stable at higher temperatures than ICM1 or ICM2.

One puzzling feature is that this model treats only an $xy$-type interaction, while the isotropic nature of the europium spins suggests a Heisenberg-like isotropic interaction. Several mechanisms exist to generate an anisotropic behavior in a Heisenberg system \cite{colarieti2003origin, biswas2011estimation}, as well as the inclusion of a dipole interaction. It is possible that the $z$ component is only slightly smaller than the $x$ and $y$ components, and thus does not need to be considered in this phenomenological model, as it does not affect the energies of the actual ground states. Indeed, the results for out-of-plane field do not change substantially for a weak (nearly Heisenberg) anisotropy, but would likely change the behavior of the model under the application of in-plane field. As such, future measurements of the in-plane phases are crucial to elucidate the degree of deviation from an isotropic model in more detail. Finally, we note that the slight easy-plane anisotropy, characterized by a saturation field of 3.2 T along the $c$ axis and 1.1 T along the $a$ axis, drives a preference for cycloidal and vortex order. Tuning of the anisotropy (perhaps by chemical doping or uniaxial strain) to be easy-axis may prefer the formation of skyrmions.
% difference in the Zeeman energy, $\mu_\text{B}gSB$, to field polarize the system is only 850 \textmu eV, much smaller than the 200 \textmu eV energy scale of the Heisenberg interactions. 

\section{Conclusion}
While \cite{kurumaji2025electronic} established the existence of multiple incommensurate phases in \ce{EuAg4Sb2}, here we fully resolve their symmetry, domain structure, and microscopic spin texture via polarized SANS, SNP, and cooling with symmetry lowering strain or magnetic fields, enabling the first complete phase-by-phase mapping and modeling for this material class.
% Herein we have elucidated the domain structure and real-space structure of the three zero field phases in \ce{EuAg4Sb2} through cooling with symmetry lowering strain or magnetic fields and polarized neutron scattering.
The variety of phases present is suggestive of a complex energy landscape hosting many phases of similar energy. The proximity of such varied phases hints at potential design principles for tunable spin moir\'{e} superlattices. Further, the demonstration of the coupling of multi-$q$ magnetic domains to strain represents a potentially fruitful route to magnetic domain control. Probing the electronic response to strain-field domain control is of interest for future investigation. The tunable vortex lattice phases present in \ce{EuAg4Sb2} therefore offer an ideal platform to study tunable topological textures in a spin moir\'{e} superlattice system.

We also note that the system displays an easy-plane anisotropy, where all three phases prefer modulated textures with spin components predominantly within the $ab$-plane. This is consistent with the easy-plane anisotropy and lower in-plane saturation field observed in magnetization measurements. If the easy-plane anisotropy is lifted, instead being isotropic or more easy-axis, the system may instead favor a skyrmion or meron-antimeron lattice state. This suggests that probing the bulk magnetic anisotropy can provide insight into what phases may be present in a frustrated rare earth magnet phase diagram before the application of more specialized techniques to determine the precise phases. This motivates targeted experiments such as in-plane field experiments and chemical or physical strain tuning of interlayer thickness which may tune the system into a phase with non-zero skyrmion number. Exploring the tunability of the phase diagram would further test our model parameters, providing a predictive loop between experiment and theory.

Further, all three phases contain fewer sites which deviate significantly from a local saturated moment than a single-$q$ transverse spin modulation. The ground state ICM1 completely conserves the saturated moment, and ICM2 and ICM3 only have the spin go to zero at point-like (anti-)vortex core centers, whereas a transverse spin modulation contains lines where the spin goes to zero. At higher temperatures in ICM2 and ICM3, the vortex and anti-vortex cores may be thermally fluctuating, while at low temperature and high field they may become field polarized, explaining the small $2q_1$-type peaks observed (see supplemental information). These observations provide insight into the design principles of multi-$q$ materials: the magnetic moment of Eu$^{2+}$ is relatively isotropic (being a half-filled $f$-shell where the electronic lobes are largely averaged out), so the moment is largely Heisenberg-like and favors any spin modulation which conserves the saturated moment. The addition of frustration and conduction electron-mediated interactions further leads to a plethora of tunable spin-moir\'{e} superlattice states.

\section{Methods}
% 4.71109, 24.62790
\subsection{Small angle neutron scattering.}
Small angle neutron scattering (SANS) was performed at the Institut Laue-Langevin on the D33 beamline, at the High Flux Isotope Reactor on the GP-SANS beamline at Oak Ridge National Laboratory, and the Swiss Spallation Neutron Source on the SANS-I beamline at the Paul Scherrer Institut. Neutrons of wavelength 3.1 to 5 \AA\ were used. The sample was rocked about the vertical and/or horizontal axes to map a 3D volume of reciprocal space. SANS measurements were analyzed in GRASP \cite{dewhurst2023graphical} and in the GRIP module \cite{neves2024grasp}. Every measurement has a high temperature background subtracted.

\subsection{Strain SANS.}
For the uniaxial strain SANS experiment, measurements were performed on SANS-I at PSI using a Razorbill UC220T piezoelectrically driven strain cell with clearance for a transmission scattering measurement on a bespoke sample stick. The strain cell is equipped with calibrated capacitive sensors which measure the displacement and force applied to the crystal, and wires for simultaneous electrical transport were included to allow \textit{in-situ} strain and transport during a SANS experiment. 
%The cell is depicted in Fig. \ref{strain_cell}\textbf{a,b}, with the force as a function of displacement plotted in Fig. \ref{strain_cell}\textbf{c} and the field-symmetrized longitudinal conductivity as a function of compression plotted in Fig. \ref{strain_cell}\textbf{d}.
This cell can apply up to 50 N or 25 \textmu m of strain (depending on the Young's modulus and length of the sample), with a standard sample size of $\sim0.1\times1\times3$ mm$^3$. 3.1 \AA\ neutrons were used to reduce beam absorption by the europium. A small portion of sample in the titanium strain cell jaws was relatively unstrained; small spot-size measurements confirmed that less domain imbalance was present in this region. Unpolarized SANS measurements were analyzed in GRASP \cite{dewhurst2023graphical} and in the GRIP module \cite{neves2024grasp}.

\subsection{Polarized SANS.}
Polarization analysis (PA) SANS \cite{honecker2013theory, muhlbauer2019magnetic, dewhurst2023graphical} was performed on the D33 beamline at the Institut Laue–Langevin in Grenoble, France. The incident beam is spin-polarized along the vertical axis by an FeSi supermirror polarizer, and can be flipped with an RF spin flipper. The outgoing neutron spin is measured with a cell of gaseous $^{3}$He where the nuclear spins are polarized to $\sim78\%$. The neutrons are selectively absorbed by the $^{3}$He when their spin is anti-aligned with the $^{3}$He nuclear spin \cite{andersen2005first, batz20053he}. The polarization of the $^{3}$He is maintained with a constant field chamber and can be flipped through the application of an RF pulse in a so-called ``magic box.'' In this experiment, one quadrant of the SANS pattern was fully covered by the $^{3}$He allowing access to all symmetry-distinct peaks. Additionally, the polarization efficiency was monitored by regularly measuring the intensity of the direct beam in both the spin-flip and non-spin-flip channels throughout the course of the experiment to characterize the depolarization of the $^{3}$He cell and allow for correction of the imperfect polarization. The scattering intensity was then corrected for non-magnetic background and an imperfect flipping-ratio to the true polarization-dependent scattering cross-sections $\frac{d\sigma^{\pm\pm}}{d\Omega}$ and $\frac{d\sigma^{\pm\mp}}{d\Omega}$ \cite{wildes2006scientific, dewhurst2023graphical}.

4.7 \AA\ neutrons were used, and the $^{3}$He cell was filled to 0.8 bar with a 10 cm cell length resulting in an estimated transmission of 28\% and an estimated filtering efficiency of 97\%. This resulted in a total polarization-analysis efficiency starting at $\sim90\%$ which decayed with a time constant of $\sim 200$ hrs. In this geometry, the polarization dependent magnetic scattering (neglecting all nuclear and nuclear-magnetic terms) is given by
\begin{equation}
    \frac{d\sigma^{\pm\pm}}{d\Omega} \propto |M_{\perp, z}^2|
\end{equation}
and
\begin{equation}
    \frac{d\sigma^{\pm\mp}}{d\Omega} \propto |M_{\perp, x}^2| + |M_{\perp, y}^2| \mp i(M_{\perp, x}M_{\perp, y}^*-M_{\perp, x}M_{\perp, y}^*).
\end{equation}
$\bm{M}_\perp(\bm{Q})=\hat{\bm{Q}}\times\{\bm{M}(\bm{Q})\times\hat{\bm{Q}}\}$ is the component of $\bm{M}(\bm{Q})$ perpendicular to $\bm{Q}$, $x$, $y$, and $z$ are the conventional polarization coordinates indicated in Fig. \ref{fig:pSANS}g. Hence, the SF channel scattering quantifies the component of modulated magnetization in the plane perpendicular to $\bm{Q}$ (a transverse in plane spin modulation), and the NSF channel quantifies the component of modulated magnetization out of the plane along $\bm{c}$. Additionally, the difference between $\frac{d\sigma^{+-}}{d\Omega}$ and $\frac{d\sigma^{-+}}{d\Omega}$ quantifies the chiral term which measures to what degree $M_{\perp, x}$ and $M_{\perp, y}$ are out of phase with one another. Additional details of the polarization analysis may be found in Supplementary Information.

\subsection{Spherical Neutron Polarimetry.}
Due to the relatively small wave vector of the ICM phases, a cold neutron instrument is ideal to observe the magnetic peaks in \ce{EuAg4Sb2}. Therefore, we employed the cold triple-axis spectrometer ThALES at the Institut Laue–Langevin in Grenoble, France which can be equipped with the Cryopad (CRYOgenic Polarization Analysis Device) spherical neutron polarimeter \cite{tasset1999spherical}. Analysis and visualization of SNP data was performed with Mag2Pol \cite{qureshi2019mag2pol}. Both statistical and estimated systematic uncertainties (following \cite{giles2020imitation}) were included. The sample was manually reloaded multiple times with different $\psi$ (the angle between the scattering plane and the $a^*$ axis) to place the different magnetic peaks of interest in the horizontal scattering plane of the instrument: $\psi=0^\circ$ for the $(H0L)$ peaks of ICM1 and ICM3, $\psi\approx13.5^\circ$ for the ICM2 peaks, and $\psi=30^\circ$ for the $(HHL)$ ICM3 peaks. 5 Å neutrons and a beam collimator were employed to provide sufficient resolution to observe the magnetic peaks of interest. The principle peaks in the zeroth zone appear at a $2\theta$ of 7.3$^\circ$ from the direct beam, which is a large enough angle to measure reliably despite the elevated background from the direct beam. In this configuration, the magnetic satellites can be observed about the (000), (003), and (006) peaks for all $\psi$, and the (101), (102), (104), and (105) peaks can be accessed for the $\psi=0$ measurement in the $(H0L)$ scattering plane. The satellite peaks observed about the (003) and (006) sum together the propagation vector from both crystallographic twins due to the broader instrument resolution along $\bm{Q}$. Additional details of the polarization analysis may be found in Supplementary Information.

\subsection{Phenomenological Model.}
For the momentum-resolved interaction in Eq.~(2), we adopt the following wave vectors: 
$\bm{Q}_a= Q(1,0)$, $\bm{Q}_b= Q(-1/2,\sqrt{3}/2)$, $\bm{Q}_c= Q(-1/2,-\sqrt{3}/2)$, $\bm{Q}_d=(Q, Q')$, $\bm{Q}_e=(-Q/2-\sqrt{3}Q'/2, \sqrt{3}Q/2-Q'/2)$, $\bm{Q}_f=(-Q/2+\sqrt{3}Q'/2, -\sqrt{3}Q/2-Q'/2)$, $\bm{Q}_g=(Q, -Q')$, $\bm{Q}_h=(-Q/2+\sqrt{3}Q'/2, \sqrt{3}Q/2+Q'/2)$, $\bm{Q}_i=(-Q/2-\sqrt{3}Q'/2, -\sqrt{3}Q/2+Q'/2)$, $\bm{Q}_j= Q''(0, 1)$, $\bm{Q}_k= Q''(-\sqrt{3}/2, -1/2)$, and $\bm{Q}_l= Q''(\sqrt{3}/2, -1/2)$ with $Q=\pi/3$, $Q'=\sqrt{3}\pi/18$, and $Q''=\sqrt{3}\pi/6$, where we consider the interaction parameters $G_1=1$ and $A_1=0.05$ for $(\bm{Q}_a, \bm{Q}_b, \bm{Q}_c)$, $G_2=0.95$ and $A_2=0.1$ for $(\bm{Q}_d, \bm{Q}_e, \bm{Q}_f, \bm{Q}_g, \bm{Q}_h, \bm{Q}_i)$, and $G_3=0.9$ and $A_3=0.16$ for $(\bm{Q}_j, \bm{Q}_k, \bm{Q}_l)$. 

The optimized spin configurations for the model are obtained through an iterative simulated annealing scheme combined with single-spin updates based on the Metropolis Monte Carlo algorithm. 
Periodic boundary conditions were imposed throughout, and simulations were performed for multiple lattice sizes to verify the robustness and convergence of the numerical procedure.

\section{Acknowledgements}
We appreciate fruitful discussions with C. John and S. Moody.   This work was funded, in part, by the Gordon and Betty Moore Foundation EPiQS Initiative, grant no. GBMF9070 to J.G.C. (instrumentation development); the US Department of Energy (DOE) Office of Science, Basic Energy Sciences, under award no. DE-SC0022028 (material development); the Office of Naval Research (ONR) under award no. N00014-21-1-2591 (advanced characterization); and the Air Force Office of Scientific Research (AFOSR) under award no. FA9550-22-1-0432 (magnetic structure analysis).  Strain method development was supported through the Center for the Advancement of Topological Semimetals (CATS), an Energy Frontier Research Center (EFRC) funded by the US Department of Energy (DOE) Office of Science, through the Ames National Laboratory under contract DE-AC0207CH11358.  S. Hayami was supported by JSPS KAKENHI (JP23H04869), JST CREST (JPMJCR23O4), and JST FOREST (JPMJFR2366) (phenomenological modeling). J.S.W. acknowledges financial support from the Laboratory for Neutron Scattering and Imaging at PSI for an extended research visit of P.M.N., as well as funding from the Swiss National Science Foundation (SNF) under grant no. 200021$\textunderscore$188707. This work is based partly on experiments performed at the Swiss
spallation neutron source SINQ, Paul Scherrer Institute, Villigen, Switzerland. A portion of this research used resources at the High Flux Isotope Reactor, a DOE Office of Science User Facility operated by the Oak Ridge National Laboratory. The beam time was allocated to GP-SANS on proposal number IPTS-32248.1. 
ILL polarized neutron scattering experiments may be accessed at http://doi.ill.fr/10.5291/ILL-DATA.DIR-324, https://doi.ill.fr/10.5291/ILL-DATA.5-54-408, and https://doi.ill.fr/10.5291/ILL-DATA.5-71-2

\subsection{Contributions}
Small angle neutron scattering was performed by P.M.N. and J.P.W. with support from R.C., L.M.D.-S., and J.S.W.. Strain experiments were performed by P.M.N. with support from J.C.P., M.B., M.Z., and J.S.W.. Spherical neutron polarimetry was performed by P.M.N. with support from A. H., P. S., and N. Q.. Materials were synthesized and characterized by T. K.. Phenomenological modeling was performed by S.H.. All authors contributed to writing the manuscript. J.S.W. and J.G.C. coordinated the project.

% \bibliography{references}


\end{document}